\documentclass[10pt,english]{article}
\usepackage{subfigure}
\usepackage{graphicx}
\usepackage{float}
\usepackage[T1]{fontenc}
\usepackage[latin9]{inputenc}
\usepackage[margin=1in]{geometry}
\usepackage{float}
\usepackage{amsmath}
\usepackage{amsbsy}
\usepackage{setspace}
\usepackage{amssymb}
\usepackage{esint}
\usepackage{cite}
\usepackage{hyperref}
\newfloat{algorithm}{tbp}{loa}
\floatname{algorithm}{Algorithm}
\usepackage{algorithmic}
\usepackage{hyperref}

\newlength\myindent
\makeatletter

\floatstyle{ruled}
\newfloat{algorithm}{tbp}{loa}
\floatname{algorithm}{Algorithm}

\usepackage{babel}

\begin{document}

\title{Sandbox Sample Classification Using \\ Behavioral Indicators of Compromise}

\author{M. Andrecut}

\date{January 17, 2020}

\maketitle
{

\centering Calgary, Alberta, Canada

\centering mircea.andrecut@gmail.com

} 

\bigskip

\bigskip

\begin{abstract}

Behavioral Indicators of Compromise are associated with various automated methods used to extract the sample behavior by observing the system function calls performed in a virtual execution environment. Thus, every sample is described by a set of BICs triggered by the sample behavior in the sandbox environment. Here we discuss a Machine Learning approach to the classification of the sandbox samples as MALICIOUS or BENIGN, based on the list of triggered BICs. 
Besides the more traditional methods like Logistic Regression and Naive Bayes Classification we also discuss a different approach inspired by the statistical Monte Carlo methods. 
The numerical results are illustrated using ThreatGRID and ReversingLabs data. 

\bigskip


\end{abstract}

\bigskip

\section{Introduction}

The amount of malware distributed via Internet is increasing every year, creating difficult challenges for security analysts. Hundred of millions of malware samples are identified every year, requiring an increased 
analysis effort. An automatic classification approach based on Machine Learning (ML) may provide the analysts with the necessary tools to cope with the constantly evolving malware landscape. 

Behavioral Indicators of Compromise (BIC) are associated with various automated methods used to extract the sample behavior by observing the system function calls performed in a virtual execution environment. Thus, every sample is described by a set of BICs triggered by the sample behavior in the sandbox environment. The main requirement is to classify the sandbox samples as MALICIOUS or BENIGN, based on the list of triggered BICs. 
There are many methods for ML classification problems described in the literature, however we could not identify any paper where ML is successfully applied to the MALICIOUS or BENIGN sample classification problem as stated above. 

Most of the literature papers are concerned with the classification of the type of malware (see the references \cite{key-1}-\cite{key-9}). 
For example, the problem described in [9] is to identify the family class of malware from several considered classes. 

We should stress here that the problem we intend to solve is different than the type of malware classification problem, namely the problem we intend to solve here is to classify the sandbox samples as MALICIOUS or BENIGN. 
Moreover, the classification targets, that is the type of data of these two problems are completely different. For example, the type of data used in [9] consists of malware samples with 31 features, and the results are obtained from 7,009 samples. The data used in our problem is several order of magnitudes larger, that is the number of sandbox samples considered span a full year, 163 million samples from ThreatGRID and ReversingLabs, and the samples are described by 1,817 BICs, which are different than the 31 features considered in [9]. 

Besides the more traditional methods like Logistic Regression (LR) and Naive Bayes (NB) Classification we also discuss a different approach inspired by the statistical Monte Carlo methods, which we called the Max Precision (MP) method. We show that for this particular problem and data, the MP method gives better results than the LR and NB methods.

\section{Data Description}

The sandbox data considered here consists of one year of CISCO (\url{https://www.cisco.com}) ThreatGRID (TG) data, starting on February 1, 2019, and ending on January 31, 2020. The data is organized in daily files, containing all the samples from that particular day. 
In total there are 163,493,486 unique samples. Each data sample is identified by a unique hash string (sha256). By using these unique hash values one can query a third party source like ReversingLabs, in order to obtain a classification of the samples. 

ReversingLabs TitaniumCloud Reputation Services (\url{https://www.reversinglabs.com}) is a threat intelligence solution with up-to-date threat classification and rich context on over 8 billion goodware and malware files. ReversingLabs does not depend on crowdsourced collection but instead curates the harvesting of files from multiple anti-virus vendors and diverse malware intelligence sources. 
Using the unique hash and the ReversingLabs API all the TG samples have been classified using the industry reputation consensus. Thus, for each TG sample we could extract the following classification values from ReversingLabs: MALICIOUS, KNOWN, SUSPICIOUS, UNKNOWN. Finally, only the samples classified as MALICIOUS or KNOWN have been used in training and testing the classification methods discussed here. 
In order to simplify the notation and to be consistent with our problem statement, we assume that the KNOWN samples are considered BENIGN. In total we identified 32,627,101 MALICIOUS samples and 68,501,921 BENIGN (or KNOWN) samples. That is an average of 89,389 MALICIOUS and 187,676 BENIGN samples per day. 

Each sample is characterized by a number of 1,817 BICs. Thus, each sample $X_n$ can be described by $M=1817$ dimensional binary vectors: 
\begin{equation}
X_n = [x_{n,0},x_{n,1},...,x_{n,M-1}]^T, \quad n=0,1,...
\end{equation}
where $T$ is the vector transpose, and the components $x_{n,m} \in\{0,1\} $, $m=0,1,...,M-1$, are binary variables associated with the BICs. 
That is, $x_{n,m}=1$ if BIC $m$ was triggered by the sample $X_n$, and $x_{n,m}=0$ otherwise. 
Also, each sample $X_n$ has an assigned classification label, MALICIOUS or BENIGN (obtained from ReversingLabs, as described above):
\begin{equation}
Y_n \in \{0,1\}, \quad n=0,1,...
\end{equation}
that is $Y_n=1$ if the sample $X_n$ is MALCIOUS, and $Y_n=0$ if it is considered BENIGN (KNOWN). 

In order to train and test our models we assume that the data from the current day is not yet classified, 
and we only know the classification of the data from the previous days. This way we aim to classify the data from the current day, given the classified data from the previous days. 
That is, we have a loop with the index corresponding to the current day, and we consider the historical data previous to the current day of the loop, and with this data 
we train the models and we classify all the samples from the current day (which are considered new, and have not been used in the training process), 
we compute the evaluation metrics, and then we advance in the loop to the next day. 

\section{Logistic Regression Classifier}

Since the TG data considered here has a binary representation, it is natural to first consider the LR, which is a binary classification method frequently used in ML \cite{key-10}). 

The LR classification model is a binary classification model in which the conditional probability of one of the two possible realizations of the output variable (0 or 1) is assumed to be equal to a linear 
combination of the input variables, transformed by the logistic function.

Consider that $(X_n,Y_n)$, $n=0,1,...,N-1$ is the training data. The probability that the output $Y_n$ is equal to 1, conditional on the input $X_n$, is assumed to be: 
\begin{equation}
p(Y_n=1 \mid X_n) = S(W^TX_n),
\end{equation}
where 
\begin{equation}
S(z) = \frac{1}{1+\exp (-z)},
\end{equation}
is the logistic function, and $W=[w_0,w_1,...,w_{M-1}]^T$ is an $M-$ dimensional vector of coefficients (weights) to be determined during the learning process. 

One can easily see that the logistic function has the following properties:
\begin{align}
\lim_{z\rightarrow -\infty} S(z) &= 0, \\
\lim_{z\rightarrow +\infty} S(z) &= 1, \\
0 < S(z) &< 1,
\end{align}
and therefore since the probabilities need to sum up to 1, the probability that $Y_n=0$, conditional on the input $X_n$, is:
\begin{equation}
p(Y_n=0 \mid X_n) = 1 - p(Y_n=1 \mid X_n) = 1 - S(W^TX_n).
\end{equation}

Once the coefficient vector $W$ is learned using the training data, a new sample is classified as following:
\begin{equation}
\text{class}_{LR} (X) =   \begin{cases}
   1\quad (\text{MALICIOUS}) & if \quad p(Y=1 \mid X) \geq 0.5 \\
   0\quad (\text{BENIGN}) & \text{otherwise}
  \end{cases}.
\end{equation}

The estimation of the coefficient vector $W$ is done using the Maximum Likelihood method, that is:
\begin{equation}
W = \max_W \lbrace L(W,Y,X) = \prod_{n=0}^{N-1} p(Y=Y_n \mid X=X_n) \rbrace,
\end{equation}
or equivalently, the minimization of the negative log-likelihood:
\begin{equation}
W = \min_W \lbrace \mathcal{L}(W,Y,X) = -\log L(W,Y,X) \rbrace,
\end{equation}

Using the logistic function, the negative log-likelihood is given by:
\begin{align}
\mathcal{L}(W,Y,X) = - \sum_{n=0}^{N-1} \lbrace Y_n (W^TX_n) - \log(1+e^{W^TX_n}) \rbrace.
\end{align}

One can easily show that minimum of the negative log-likelihood estimator $\mathcal{L}(W,Y,X)$ (when it exists) can be obtained using the Newton-Raphson algorithm as follows: 
\begin{equation}
W^{new} = W^{old} - \left[ \frac{\partial^2 \mathcal{L}(W,Y,X)}{\partial W \partial W^T} \right]^{-1} \frac{\partial \mathcal{L}(W,Y,X)}{\partial W} , 
\end{equation}
where:
\begin{equation}
\frac{\partial \mathcal{L}(W,Y,X)}{\partial W} = -\sum_{n=0}^{N-1} X_n[Y_n - S(W^TX_n)],
\end{equation}
and respectively:
\begin{equation}
\frac{\partial^2 \mathcal{L}(W,Y,X)}{\partial W \partial W^T} = \sum_{n=0}^{N-1} X_n X_n^T S(W^TX_n)[1 - S(W^TX_n)].
\end{equation}

\section{Naive Bayes Classifier}

The Naive Bayes (NB) classifier is another suitable ML method to be used for TG data classification. 
Comparing to the LR method (and other ML methods), the NB classifier has the advantage that the maximum-likelihood training can be done in linear time by evaluating a closed-form expression, 
rather than using expensive iterative approximation algorithms. 

The NB classifier is a conditional probability model \cite{key-10}. We describe the model in the more general case of $K$ classes, $C=\{0,1,...,K-1\}$, and then we particularize it to the binary classification 
problem, that is when $K=2$, and $C=\{0,1\}$, with BENIGN=0 and MALICIOUS=1. 

Consider that $(X_n,Y_n)$, $n=0,1,...,N-1$ is the training data, such that $Y_n \in C = \{0,1,...,K-1\}$.
Given a new sample $X = [x_0,x_1,...,x_{M-1}]^T$ to be classified, the NB model estimates the probabilities:
\begin{equation}
p(Y=k \mid X), \quad k \in C,
\end{equation}
and the sample $X$ is classified as $Y = k$, where $k$ is the index of the class with maximum probability value:
\begin{equation}
\text{class}_{NB}(X) = k = \text{arg} \max_{k \in C} p(Y=k \mid X).
\end{equation}

Using Bayes' theorem the conditional probabilities can be written as:
\begin{equation}
p(Y=k \mid X) = \frac{p(Y=k)p(X \mid Y=k)}{p(X)} = \frac{\text{prior} \times \text{likelihood}}{\text{evidence}} = \text{posterior}, \quad k \in C.
\end{equation}
The "naive" assumption of the NB classifier is that all the $M$ features of the sample $X=[x_0,x_1,...,x_{M-1}]^T$ are mutually independent conditional on the class $k$, that is we can write 
the likelihood as a product of independent probabilities:
\begin{equation}
p(X \mid Y=k) = \prod_{m=0}^{M-1} p(x_m|Y=k), \quad k \in C.
\end{equation}
With these assumptions the $K$ posterior conditional probabilities are:
\begin{equation}
p(Y=k \mid X) = \frac{1}{Z} p(Y=k) \prod_{m=0}^{M-1} p(x_m|Y=k), \quad k \in C,
\end{equation}
where $Z$ is a normalization constant, independent on class $k$ and it can be discarded. 

If the number of classes is $K=2$ and the features are also modeled as binary variables, which is exactly our case, then the problem can be simplified further by assuming that the features of $X$ are modelled as 
Bernoulli random variables (independent Boolean random variables). In this case the likelihood can be estimated as:
\begin{equation}
p(X \mid Y=k) = \prod_{m=0}^{M-1} p_{km}^{x_m}(1-p_{km})^{1-x_m},
\end{equation}
where $p_{km}$ is the probability of class $k \in C$ generating the feature $x_m$. 

Using $N$ training samples $X_n$, one can show that:
\begin{equation}
p_{km}=\frac{n_{km}}{N_k}, \quad k \in C, \quad m=0,1,...,M-1,
\end{equation}
where $n_{km}$ is the number of samples from class $k$ that contain the particular feature (BIC) $x_m$, and $N_k$ is the number of samples from the class $k$. 

During training we may encounter a feature that shows up in one class, but not in the other. In this case, the log sum of likelihood would diverge to negative infinity. In order to cope 
with this problem we use the Laplace smoothing, which assumes an initial likelihood of 1/2 for each class, regardless if the feature $x_m$ is present or not, and therefore: 
\begin{equation}
p_{km}=\frac{n_{km}+1}{N_k+2}, \quad k \in C, \quad m=0,1,...,M-1.
\end{equation}

One can also show that the prior probabilities $p(Y=k)$ for each class $k \in \{0,1\}$ can be estimated as the fraction of the training samples for a given class:
\begin{equation}
p(Y=k) = \frac{N_k}{N}.
\end{equation}

Once these probabilities are calculated from the training data, a new sample $X = [x_0,x_1,...,x_{M-1}]^T$ can be classified as $Y=k$, where $k$ is the index of the class with maximum score value:
\begin{equation}
\text{class}_{NB}(X) = k = \text{arg} \max_{k \in C} s_k(X),
\end{equation}
where the score is:
\begin{equation}
s_k(X) = \log{\frac{N_k}{N}} + \sum_{m=0}^{M-1}[ x_m log \frac{n_{km}+1}{N_k+2} + (1-x_m) \log(1-\frac{n_{km}+1}{N_k+2})].
\end{equation}

\section{Maximum Precision Method}

Here we consider a heuristic method based on the statistical estimation of the precision of BICs, which we called the Max Precision (MP) method. 
This approach is inspired by the Monte Carlo statistical methods.  
In the MP method the precision is used to estimate a "score" $S$ associated with each considered BIC. For each BIC, the score is computed in two steps: 
(1) a "predictive step", where the score is initialized with the BIC precision, and (2) a "corrective step", where the score is adjusted using a penalty method 
based on BIC's classification performance on the training data. 

\subsection{Predicting Step}

With each BIC we associate a score $S_m$, $m=0,1,...,M-1$. 
Since the historical data has been already classified as MALICIOUS=1 or BENIGN=0, we initialize the score value of each BIC with the following quantity:
\begin{equation}
S_m = \frac{N^{(1)}_m}{N^{(1)}_m + N^{(0)}_m}, \quad m=0,1,...,M-1,
\end{equation}
where $N^{(1)}_m$ is the number of times the BIC $m$ has been triggered by MALICIOUS samples, and respectively $N^{(0)}_m$ is the number of times the BIC $m$ has been been triggered by BENIGN samples. 
Assuming that we have $N$ training samples $X_n$ classifid as $Y_n \in C=\{0,1\}$ then:
\begin{align}
N^{(1)}_m &= \sum_{n=0}^{N-1} Y_n x_{nm}, \\
N^{(0)}_m &= \sum_{n=0}^{N-1} (1-Y_n) x_{nm}.
\end{align}

The BICs can be also interpreted as classifiers themselves, that is we assume that there are $M$ classifiers, which are used to classify a training sample $X_n=[x_{n,0},x_{n,1},...,x_{n,M-1}]$ with the output values $x_{n,m} \in \{0,1\}$, $m=0,1,...,M-1$. If $x_{n,m}=1$ then the BIC $m$ has classified the sample $X_n$ as MALICIOUS=1, otherwise as BENIGN=0.

Therefore, the number $N^{(1)}_m$ is the number $t_m$ of True Positive (TP), and $N^{(0)}_m$ is the number $f_m$ of False Positive (FP) classifications performed by the BIC classifier $m$ on the training samples. 
Thus, the initial value of the score is the precision of the BIC classifiers applied to the training data: 
\begin{equation}
S_m = \frac{t_m}{t_m + f_m} = P_m, \quad m=0,1,...,M-1.
\end{equation}

Assuming that we are given a new sample $X=[x_{0},x_{1},...,x_{M-1}]$, then we can find the BIC with the highest score value:
\begin{equation}
m = \arg \max_{m=0,1,...,M-1} x_m S_m,
\end{equation}
and we classify the sample as following:
\begin{equation}
\text{class}_{MP}(X,T,m)=\begin{cases}
   1\quad (\text{MALICIOUS}) & if \quad S_{m} \geq T \\
   0\quad (\text{BENIGN}) & \text{otherwise}
  \end{cases},
\end{equation}
where $0<T<1$ is a threshold value, which can be easily determined (by searching on a grid, for example) such that the sum of classification 
errors made on the training data set is minimized:
\begin{equation}
T = \arg \min_{0<T<1} \sum_{n=0}^{N-1} [1-\delta(Y_n,\text{class}_{MP}(X_n,T))]. 
\end{equation}
Here, $\delta(a,b)$ is the Kronecker function:
\begin{equation}
\delta(a,b)=\begin{cases}
   1 & if \quad a=b \\
   0 & \text{otherwise}
  \end{cases}. 
\end{equation}

\subsection{Corrective Step}

The MP method works quite well using just the prediction step. However, better results can be obtained by 
employing a corrective step which is meant to penalize the misclassifications on the training data and to adjust the score values of the BICs. 

The corrective step consists of applying the MP classification to the training data set $(X_n,Y_n)$, $n=0,1,...,N-1$, and every 
misclassification is used as a penalty to update the score values. The corrective algorithm can be summarized as following:

For $n=0,1,...,N-1$ do:

- find the index of the BIC with the highest score for the training sample $X_n$:
\begin{equation}
m = \arg \max_{m} x_{n,m} S_m
\end{equation}

- classify training sample $X_n$: 
\begin{equation}
c_n = \text{class}_{MP}(X_n,T,m)
\end{equation}

- if misclassification occurs, then apply the following penalty rules (here $:=$ is the value assignment operator):

\begin{align}
if \quad c_n = 1 \quad and \quad Y_n = 0 \quad and \quad t_m > 0 \quad then &:\\
& t_m := t_m - 1 \\
& f_m := f_m + 1 
\end{align}
\begin{align}
if \quad c_n = 0 \quad and \quad Y_n = 1 \quad and \quad f_m > 0 \quad then &:\\
& t_m := t_m + 1 \\
& f_m := f_m - 1 
\end{align}

After this corrective step, we recalculate the scores with the new values for $t_m$ and $f_m$:
\begin{equation}
S_m = \frac{t_m}{t_m + f_m} = P_m, \quad m=0,1,...,M-1.
\end{equation}
These adjusted scores are then used to classify any new samples using Eqs. (31)-(32). 

\section{Numerical Results}

In order to estimate the quality of the classification of the new samples we consider the following metrics:

- Positive Predictive Value:
\begin{equation}
PPV = \frac{tp}{tp+fp}
\end{equation}

- Negative Predictive Value:
\begin{equation}
NPV = \frac{tn}{tn+fn}
\end{equation}

- Sensitivity:
\begin{equation}
SNS = \frac{tp}{tp+fn}
\end{equation}

- Specificity:
\begin{equation}
SPC = \frac{tn}{tn+fp}
\end{equation}

- Accuracy:
\begin{equation}
ACC = \frac{tp+tn}{tp+tn+fp+fn}
\end{equation}
where $tp$, $tn$, $fp$, $fn$ are the number of samples from the new day which have been classified as True Positive (TP), True Negative (TN), False Positive (FP), False Negative (FN).  
All the above metrics take values from 0 (bad performance) to 1 (good performance). 

All three methods LR, NB and MP have been implemented Python, such that they can handle the large amount of data efficiently. 

Our numerical estimations using Eq. (33) show that $T=0.99$ is a good "rule of thumb" for the value of the classification threshold in the MP method. This is equivalent of saying that the 
sample has triggered a BIC which has a probability of 0.99 of being triggered by a MALICIOUS sample. The results presented here are obtained using $T=0.99$. 

In what follows we denote by Max Precision 1 the Max Precision method with the prediction step only, and by 
Max Precision 2 the Max Precision method with both the prediction and the correction steps. 

In Figure 2, and below, we give the average results of the metrics for all the data considered here:

- Max Precision 2: $ACC=0.982$, $PPV=0.988$, $NPV=0.978$, $SNS=0.964$, $SPC=0.993$.

- Max Precision 1: $ACC=0.977$, $PPV=0.988$, $NPV=0.971$, $SNS=0.951$, $SPC=0.993$.

- Naive Bayes Classifier: $ACC=0.966$, $PPV=0.965$, $NPV=0.967$, $SNS=0.945$, $SPC=0.979$.

- Logistic Regression: $ACC=0.932$, $PPV=0.994$, $NPV=0.903$, $SNS=0.824$, $SPC=0.997$.

\begin{figure}[!ht]
\centering \includegraphics[width=13cm]{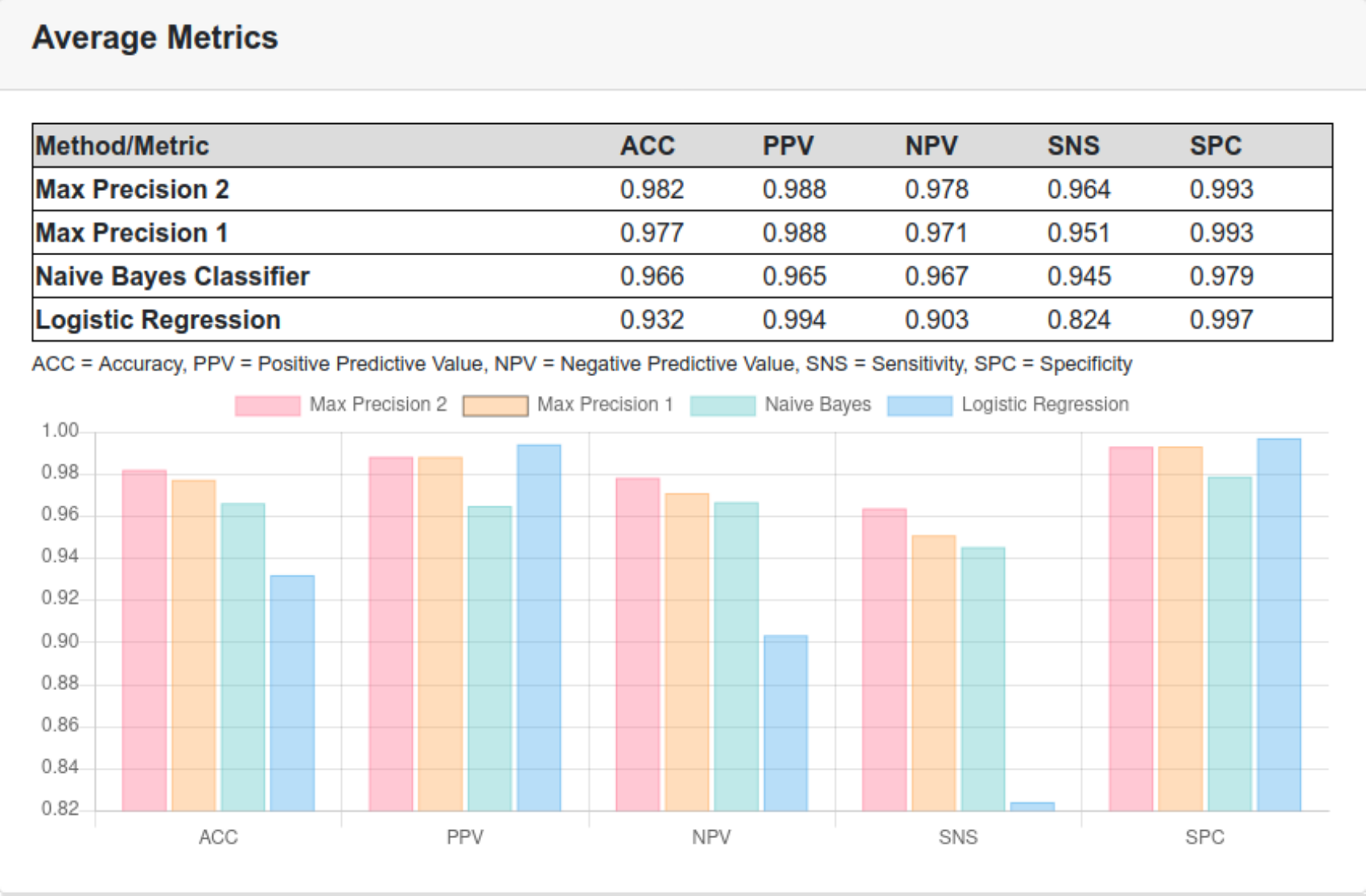}
\caption{The average metric values for the Max Precision 1 and 2 (with T=0.99), Naive Bayes Classifier and Logistic Regression heuristic methods.}
\end{figure}

In Figure 3 we give the Accuracy (ACC) chart for the Max Precision 1 and 2 (with T=0.99), Naive Bayes Classifier and Logistic Regression heuristic methods. One can see how well balanced and stable the results of the Max Precision method. 

\begin{figure}[!ht]
\centering \includegraphics[width=16.5cm]{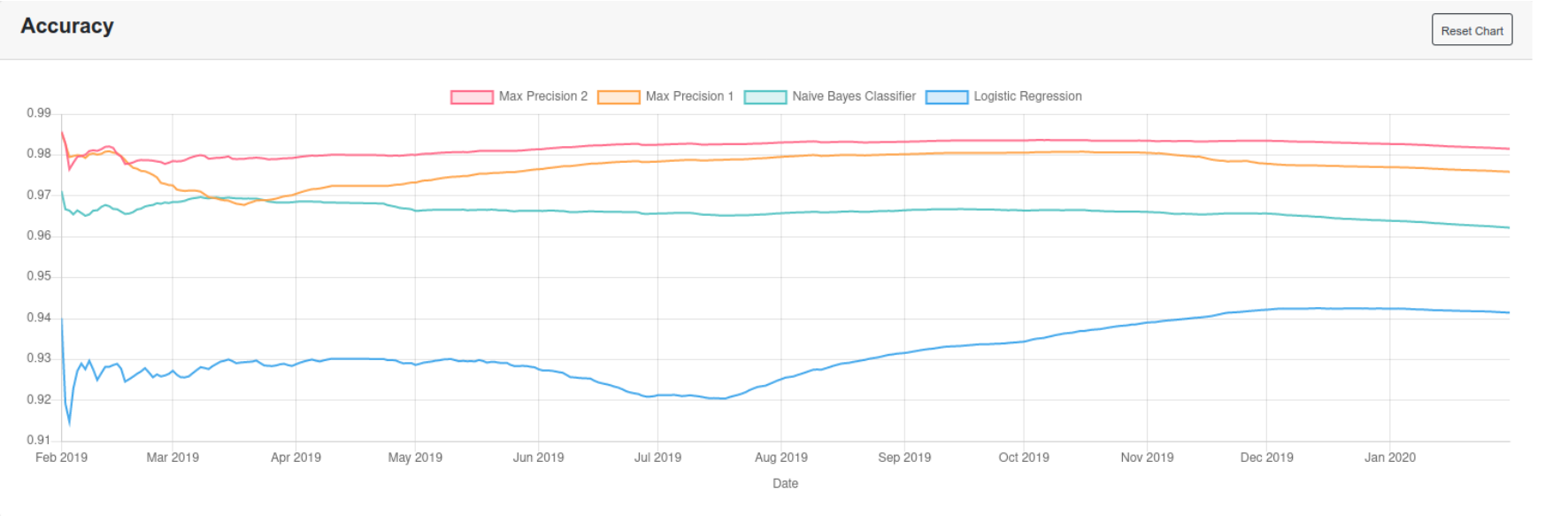}
\caption{The Accuracy values for the Max Precision 1 and 2 (with T=0.99), Naive Bayes Classifier and Logistic Regression heuristic methods.}
\end{figure}

These numerical results clearly show that the MP method gives better accuracy, and in general more stable and balanced results, than the NB and the LR methods. It is worth noticing 
that using the two predictive-corrective steps in MP2 significantly improves the results over the MP1, which only uses only the predictive step. 
Also, the training for the MP method is performed in linear time, and therefore it does not require an elaborated and lengthy training process like the other ML methods.  

\section{Conclusion}
We have discussed a ML approach to the classification of the sandbox samples as MALICIOUS or BENIGN, based on the list of triggered BICs. 
Our numerical results show that for this problem and data, the MP method which is inspired by Monte Carlo methods, and it is based on the statistical estimation of the BIs precision, outperforms the more traditional LR and NB methods.

\end{document}